\documentclass[pra,twocolumn,showpacs]{revtex4}
\usepackage{amsmath,amssymb,mathrsfs}
\usepackage{psfrag}
\usepackage{graphicx}
\usepackage{graphics}
\usepackage{epsfig}
\usepackage{bm}
\usepackage{color}
\usepackage{verbatim,color,ulem}

\newcommand{\beq}{\begin{equation}}
\newcommand{\eeq}{\end{equation}}
\newcommand{\beqa}{\begin{eqnarray}}
\newcommand{\eeqa}{\end{eqnarray}}

\begin{document}

\title{Superfluidity, Sound Velocity and Quasi Condensation\\ 
in the 2D BCS-BEC Crossover}
\author{L. Salasnich$^{1,2}$, P.A. Marchetti$^{1,3}$, and F. Toigo$^{1,2}$}
\affiliation{$^{1}$Dipartimento di Fisica e Astronomia ``Galileo Galilei'', 
Universit\`a di Padova, Via Marzolo 8, 35131 Padova, Italy
\\
$^{2}$Consorzio Nazionale Interuniversitario per le Scienze Fisiche 
della Materia (CNISM), Unit\`a di Padova, Via Marzolo 8, 35131 Padova, Italy
\\
$^{3}$Istituto Nazionale di Fisica Nucleare, Sezione di Padova, 
Via Marzolo 8, 35131 Padova, Italy}

\date{\today}

\begin{abstract}
We study finite-temperature properties of a two-dimensional superfluid 
made of ultracold alkali-metal atoms in the BCS-BEC crossover. 
We investigate the region below the critical 
temperature $T_{BKT}$ of the Berezinskii-Kosterlitz-Thouless 
phase transition, where there is quasi-condensation, 
by analyzing the effects of phase and amplitude fluctuations 
of the order parameter. In particular, we calculate the superfluid 
fraction, the sound velocity and the quasi-condensate fraction 
as a function of the temperature and of the binding 
energy of fermionic pairs. 
\end{abstract}


\maketitle

\section{Introduction}

Nowadays the manipulation of the binding energy through external magnetic 
fields (Feshbach-resonance technique) 
enables experimentalists to evolve clouds 
of two-component fermionic atoms 
from the weakly coupled BCS-like behavior of Cooper pairs to the 
strongly coupled Bose-Einstein 
condensation (BEC) of molecules \cite{exp_MolecularBEC}. This 
transition is characterized by a crossover in which 
the $s$-wave scattering length $a_s$ of the inter-atomic potential 
diverges as it changes sign 
\cite{exp_Crossover,chin}. Recently, a considerable theoretical effort 
\cite{sala-odlro,ortiz,ohashi2,sala-odlro3,sala-odlro4,
cinesi,sala-odlro5,sala-odlro6} 
has been expended on studying the condensate fraction of such a tunable 
superfluid, also in the two-dimensional (2D) case at zero temperature 
within a mean-field approach \cite{sala}. 

Quantum and thermal fluctuations play a relevant role 
in any generic 2D superfluid system \cite{mermin,hohenberg,coleman,
nagaosa,atland,stoof,randeria}. 
In the last years a beyond-mean-field formalism which takes into account 
fluctuations of the order parameter 
has been developed for 2D Fermi superfluids 
\cite{marini,loktev0,babaev,loktev,tempere,tempere2}. 
The recent experimental observation \cite{nature} of a pairing pseudogap 
in a 2D Fermi gas has strongly 
renewed the interest on this subject. 

In this paper we use this formalism to 
study the superfluid density and the sound velocity 
of the 2D Fermi superfluid as a function of the temperature and of the 
binding energy of fermionic pairs. 
The rest of the paper is organized as follows:
the finite-temperature path-integral formulation 
of the problem is discussed in Section III; 
the mean-field approach to the 2D BCS-BEC crossover 
is reported in Section III. 
The effect of fluctuations of the phase of the order parameter
is analyzed in Section IV, where the 
superfluid fraction is evaluated
as a function of the temperature for different 
values of the binding energy. In Section V we consider 
the effect of amplitude fluctuations of the order parameter 
in the determination of the sound velocity of the uniform superfluid system 
in the crossover: at zero temperature we compare the quite different 
results obtained with and without amplitude 
fluctuations (in 2D but also in 3D). 
In Section VI we calculate the quasi-condensate fraction 
of fermionic atoms in the region below the Berezinskii-Kosterlitz-Thouless 
critical temperature, where there is algebraic long-range 
order of the two-body density matrix.

\section{Formalism for fermions in two spatial dimensions}

We consider a two-dimensional Fermi gas of ultracold and dilute 
two-spin-component neutral atoms. We adopt the path integral formalism, 
where the atomic fermions 
are described by the complex Grassmann 
fields $\psi_{s} ({\bf r},\tau )$, $ \bar{\psi}_{s} ({\bf r},\tau )$ 
with spin $s = ( \uparrow , \downarrow )$  \cite{nagaosa,atland}. 
The partition function ${\cal Z}$ of the uniform 
system at temperature $T$, in a two-dimensional volume $L^2$, 
and with chemical potential $\mu$ can be written as 
\beq 
{\cal Z} = \int {\cal D}[\psi_{s},\bar{\psi}_{s}] 
\ \exp{\left\{ -{1\over \hbar} \ S  \right\} } \; , 
\eeq
where 
\beq 
S = \int_0^{\hbar\beta} 
d\tau \int_{L^2} d^2{\bf r} \ \mathscr{L}
\eeq
is the Euclidean action functional and  $\mathscr{L}$ is 
the Euclidean Lagrangian density, given by 
\beq 
\mathscr{L} = \bar{\psi}_{s} \left[ \hbar \partial_{\tau} 
- \frac{\hbar^2}{2m}\nabla^2 - \mu \right] \psi_{s} 
+ g \, \bar{\psi}_{\uparrow} \, \bar{\psi}_{\downarrow} 
\, \psi_{\downarrow} \, \psi_{\uparrow} 
\eeq
where $g$ is the strength of the s-wave inter-atomic 
coupling ($g<0$ in the BCS regime) \cite{nagaosa,atland}. 
Summation over the repeated index $s$ in the Lagrangian is meant and  
$\beta \equiv 1/(k_B T)$ with $k_B$  Boltzmann's constant. 
It is important to stress that we want to determine 
the relevant physical quantities of the system at fixed 
density $n=N/L^2$, with $N$ the total number of fermions, 
and not at fixed chemical potential $\mu$. 
For this reason we shall introduce the so-called number equation 
which enables one to express the chemical potential $\mu$ 
in terms of the density $n$. The inclusion of phase fluctuations 
in the number equation strongly modifies the functional dependence 
of $\mu$ on $n$. 

Through the usual Hubbard-Stratonovich transformation 
\cite{nagaosa,atland} the Lagrangian density $\mathscr{L}$, 
quartic in the fermionic fields, 
can be rewritten  as a quadratic form  by introducing the
auxiliary complex scalar field $\Delta({\bf r},\tau)$ so that:
\beq 
{\cal Z} = \int {\cal D}[\psi_{s},\bar{\psi}_{s}]\, 
{\cal D}[\Delta,\bar{\Delta}] \ 
\exp{\left\{ - {S_e(\psi_s, \bar{\psi_s},
\Delta,\bar{\Delta}) \over \hbar} \right\}} \; , 
\eeq
where 
\beq 
S_e(\psi_s, \bar{\psi_s},\Delta,\bar{\Delta}) = \int_0^{\hbar\beta} 
d\tau \int_{{L^2}} d^2{\bf r} \ 
\mathscr{L}_e(\psi_s, \bar{\psi_s},\Delta,\bar{\Delta})
\eeq
and the (exact) effective Euclidean Lagrangian 
density $\mathscr{L}_e(\psi_s, \bar{\psi_s},\Delta,\bar{\Delta})$ reads 
\beq 
\mathscr{L}_e =
\bar{\psi}_{s} \left[  \hbar \partial_{\tau} 
- {\hbar^2\over 2m}\nabla^2 - \mu \right] \psi_{s} 
+ \bar{\Delta} \, \psi_{\downarrow} \, \psi_{\uparrow} 
+ \Delta \bar{\psi}_{\uparrow} \, \bar{\psi}_{\downarrow} 
- {|\Delta|^2\over g} \; . 
\label{ltilde}
\eeq 
Due to to the Mermin-Wagner-Hohenberg-Coleman theorem 
\cite{mermin,hohenberg,coleman} in a 2D uniform 
system no off-diagonal long-range order (ODLRO) may exist at any finite 
temperature $T$ and this means that the critical 
temperature $T_c$ for true condensation is $T_c=0$. 
Nevertheless, below a finite temperature which is usually 
identified with the Berezinskii-Kosterlitz-Thouless critical 
temperature $T_{BKT}$ there is quasi condensation, characterized 
in our fermionic system by algebraic long-range order (ALRO) 
of the two-body density matrix, where phase fluctuations 
of $\Delta({\bf r},\tau)$ have an algebraic 
decay \cite{atland,nagaosa,stoof}. 

In this paper we want to investigate the effect of 
fluctuations of the gap field $\Delta({\bf r},t)$ around its
mean-field value $\Delta_0$ which may be taken to be real. 
For this reason we set 
\beq 
\Delta({\bf r},\tau) = \left(\Delta_0 +\sigma({\bf r},\tau) \right) 
\, e^{i \theta({\bf r},\tau)} \; , 
\label{polar}
\eeq
where $\theta({\bf r},\tau)$ is the phase of the gap field (it describes 
the Goldstone field of the U(1) symmetry) and 
$\sigma({\bf r},\tau)$ describes amplitude fluctuations.  
The adopted polar representation for $\Delta({\bf r},t)$ automatically 
satisfies Goldstone's theorem \cite{nagaosa,atland,stoof}. 

\section{Review of mean-field results}

By neglecting both phase and amplitude fluctuations, 
i.e. by setting $\theta({\bf r},t)=0$ and $\sigma({\bf r},\tau)=0$, 
and integrating over the fermionic fields 
one gets immediately the mean-field partition function 
\beq 
{\cal Z}_{mf} =  \exp{\left\{ - {S_{mf}\over \hbar} \right\}}
= \exp{\left\{ - \beta \, \Omega_{mf} \right\}} \; , 
\eeq
where 
\beqa 
{S_{mf}\over \hbar} &=& - Tr[\ln{(G_0^{-1})}] - 
\beta {L^2} {\Delta_0^2\over g} \; 
\nonumber
\\
&=& - \sum_{{\bf k}} \left[ 2
\ln{\left( 2 \cosh{(\beta E_{k}/2)} \right)} 
- \beta \Big( {\hbar^2k^2\over 2m}-\mu \Big) \right] 
\nonumber 
\\
&-& 
\beta {L^2} {\Delta_0^2\over g} \; , 
\label{omega-sp} 
\eeqa
with 
\beq 
G_0^{-1} = \left(
\begin{array}{cc}
\hbar \partial_{\tau} -{\hbar^2\over 2m}\nabla^2 -\mu & \Delta_0 \\ 
\Delta_0 & \hbar \partial_{\tau} +{\hbar^2\over 2m}\nabla^2 +\mu
\end{array}
\right)
\label{G0}
\eeq
the inverse mean-field Green function, and 
\beq 
E_k=\sqrt{\Big({\hbar^2k^2\over 2m}-\mu\Big)^2+\Delta_0^2} 
\label{ex-fermionic}
\eeq 
the energy of the fermionic elementary excitations. 
The constant, uniform and real gap parameter $\Delta_0$ can be obtained 
by minimizing  $\Omega_{mf}$ :
\beq 
{\partial \Omega_{mf}(\Delta_0) \over \partial \Delta_0} = 0 
\eeq
which gives the familiar gap equation
\beq
-{1\over g} = {1\over {L^2}} 
\sum_{\bf k} {\tanh{(\beta E_k/2)}\over 2E_k} \; . 
\eeq 
The integral on the right hand side of this equation is formally 
divergent. Nevertheless this divergence is easily removed. 
Contrary to the 3D case, in 2D a bound-state energy $\epsilon_B$ exists 
for any value of the attractive interaction strength $g$ between atoms. 
By expressing the bare interaction strength $g$ in terms of 
the physical binding energy $\epsilon_B$ through 
\cite{randeria,loktev0,babaev,tempere}
\beq 
- {1\over g} = {1\over {L^2}} \sum_{\bf k} {1\over 2{\hbar^2k^2\over 2m} 
+\epsilon_B} \; . 
\eeq
we obtain the regularized gap equation 
\beq 
\sum_{\bf k} \left( {\tanh{(\beta E_k/2)}\over 2E_k} - 
{1\over {2{\hbar^2k^2\over 2m}} +\epsilon_B} \right) = 0 \; . 
\label{gap-r}
\eeq
It is important to observe that the binding energy $\epsilon_B$ 
can be written as $\epsilon_B \simeq 2/(ma_{2D})$, where 
$a_{2D}$ is the 2D s-wave scattering length, such that 
$a_{2D} \simeq a_z \exp(−az/a_{3D})$ with $a_{3D}$ 
the 3D scattering length and $a_z$ the characteristic length 
of the strong transverse confinement which makes 
the system 2D \cite{bertaina}. 
>From Eq. (\ref{gap-r}) one obtains the energy gap $\Delta_0$ 
as a function of $T$, $\mu$, and $\epsilon_B$, 
i.e. $\Delta_0=\Delta_0(T,\mu,\epsilon_B)$. 
The total number $N$ of fermions is obtained from the familiar 
thermodynamic relation 
\beq 
N = - \left({\partial \Omega_{mf}
\over \partial \mu}\right)_{{L^2},T} \; ,  
\eeq 
which gives the number equation 
\beqa
N = \sum_{\bf k} 
\Big( 1 &-& {\hbar^2k^2/2m-\mu\over E_k} \tanh{(\beta E_k/2)} \Big) 
\label{number}
\eeqa
which must be solved together with (\ref{gap-r}) 
to determine the behavior of $\Delta_0$ and $\mu$ as a function 
of the temperature $T$ and of the binding energy $\epsilon_B$ at fixed number 
density $n=N/L^2$. At zero temperature ($T=0$) one easily finds the 
exact solutions of Eqs. (\ref{gap-r}) and (\ref{number}) as 
\beqa 
\mu &=& \epsilon_F - {1\over 2}\epsilon_B 
\; \quad \mbox{ at } T=0 \; , 
\label{ita01}
\\
\Delta_0 &=& \sqrt{2\epsilon_F\epsilon_B} \; 
\quad \mbox{ at } T=0 \; . 
\label{ita02}
\eeqa 

\begin{figure}[t]
\centerline{\epsfig{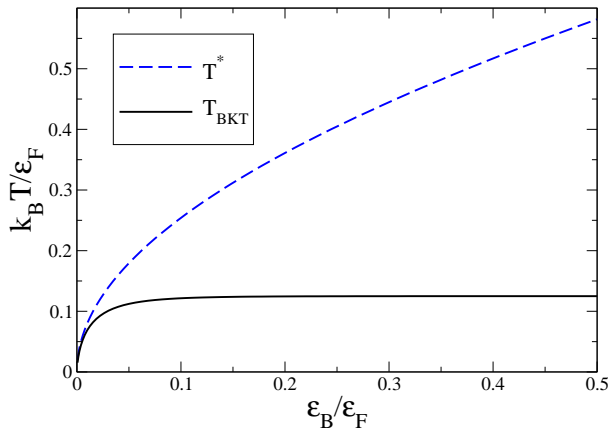}}
\small 
\caption{(Color online). Relevant temperatures of the Fermi gas 
as a function of the scaled binding energy $\epsilon_B/\epsilon_F$,  
with $\epsilon_F$ the Fermi energy. Dashed line: temperature $T^*$ above 
which the quasi-condensate $\langle |\Delta({\bf r},\tau)|\rangle$
is zero; solid line: 
Berezinskii-Kosterlitz-Thouless critical temperature $T_{BKT}$. 
Note that only at $T=0$ the condensate 
$\langle \Delta({\bf r},\tau)\rangle$ is finite.} 
\label{fig1}
\end{figure} 

We identify the temperature $T^*$ as the 
temperature at which the mean-field energy gap $\Delta_0$ 
becomes zero \cite{randeria,loktev0,babaev}. 
Thus: $\langle |\Delta({\bf r},\tau;T^*)| 
\rangle = \Delta_0(T^*)=0$. 
Setting $\Delta_0=0$ in Eqs. (\ref{gap-r}) and (\ref{number}), 
in the continuum limit $\sum_{\bf k}\to L^2\int d^2{\bf k}/(2\pi)^2$ 
and after some manipulations one obtains the equations determining $T^*$ 
as a function of $n$ (through the 2D Fermi energy $\epsilon_F=(\hbar/m)\pi n$) 
and the binding energy $\epsilon_B$:
\beqa 
\mu(T^*) = k_BT^* \; \ln{\left(e^{\epsilon_F/(k_BT^*)} - 1\right)} \; , 
\label{useit1}
\\
\epsilon_B = k_B T^* \, {\pi\over \gamma} 
\exp{\left( 
-\int_0^{\mu(T^*)/(2k_BT^*)} {\tanh{(u)}\over u} \, du \right)} \; , 
\label{useit2}
\eeqa
where $\gamma=1.781$ (see also \cite{loktev0}). 
The dashed curve shown in Fig. \ref{fig1} reports 
the scaled temperature $k_BT^*/\epsilon_F$ as a function 
of the scaled binding energy $\epsilon_B/\epsilon_F$. 
Here we limit our plot to small values of $\epsilon_B/\epsilon_F$ since
for $\epsilon_B/\epsilon_F \gtrsim 1$ and $k_BT/\epsilon_F\gtrsim 2/3$ 
beyond mean-field corrections to the number equation (\ref{number}), 
not considered above, become relevant \cite{loktev0,tempere2} 
for the determination of $T^*$ vs $\epsilon_B$ at fixed density $n$. 

Experimentally, the BCS-BEC crossover is induced 
by changing the binding energy $\epsilon_B$ with the technique of Feshbach 
resonances. As shown in Ref. \cite{sala}, the condensate fraction 
of Cooper pairs at $T=0$ is extremely small in the BCS region, 
where $\epsilon_B/\epsilon_F \ll 1$,  while it goes to one (all molecules 
are in the Bose-Einstein 
condensate) in the BEC region, where $\epsilon_B/\epsilon_F\gg 1$. 
According to Ref. \cite{sala}, for the range of scaled binding energies 
considered in Fig. \ref{fig1} the condensate fraction 
at zero temperature increases from nearly $0\%$ to about $55\%$ 
(see also Section VI). 

\section{Phase fluctuations and superfluid fraction}

We now consider the effect of phase fluctuations, i.e. in Eq. (\ref{polar})
we allow $\theta({\bf r},t)\neq 0$, but keep $\sigma({\bf r},\tau)=0$. 
To extract the contribution of the fluctuations we perform  
a gauge transformation, defining a new fermionic "neutral" field 
\beq
\chi_{s}({\bf r},\tau )= e^{i\theta({\bf r},\tau)/2}
\psi_{s}({\bf r},\tau) \; . 
\label{chi}
\eeq
In this way the Lagrangian density (\ref{ltilde}) becomes 
\beqa 
\mathscr{L}_e &=&
\bar{\chi}_{s} \left[ \hbar \partial_{\tau} 
- {\hbar^2\over 2m}\nabla^2 - \mu \right] \chi_{s}  
+ i{\hbar^2 \over 2m }\bar{\chi}_{s} {\boldsymbol \nabla} 
\theta \cdot {\boldsymbol \nabla} \chi_s 
\nonumber
\\
&+& \bar{\chi}_{s}  \chi_{s} \left[ -i {\hbar\over 2} \partial_{\tau}\theta 
- i {\hbar^2\over 4m}\nabla^2\theta 
+ {\hbar^2\over 8m}(\nabla \theta)^2 \right]
\label{lll}
\\
&+& \Delta_0 \, \chi_{\downarrow} \, \chi_{\uparrow} 
+ \Delta_0 \bar{\chi}_{\uparrow} \, \bar{\chi}_{\downarrow} 
- {\Delta_0^2\over g} \; . 
\nonumber
\eeqa 
After functional integration over the new fermionic fields 
the partition function reads \cite{nagaosa,atland}
\beq 
Z = \int {\cal D}[\theta] 
\exp{\left\{-{{\tilde S}_{e}(\theta)\over \hbar} \right\}} \; 
\eeq
where 
\beq 
{{\tilde S}_{e}(\theta) \over \hbar} = - Tr[\ln{( 
G_0^{-1} + \Sigma_{\theta} 
)}] - 
\beta {L^2} {\Delta_0^2\over g} \; 
\eeq
with $G_0^{-1}$ given by Eq. (\ref{G0}) and $\Sigma_{\theta}$ given by 
\beqa 
\Sigma_{\theta} &=& {\hat I} \left( 
i{\hbar^2\over 4m} \nabla^2\theta + i{\hbar^2\over 2m}
{\boldsymbol \nabla}\theta \cdot 
{\boldsymbol \nabla} \right) 
\nonumber
\\
&-& {\hat \tau}_3 \left( i{\hbar\over 2} \partial_{\tau} \theta  
-{\hbar^2\over 8m}(\nabla\theta)^2 \right) \; . 
\eeqa
Here ${\hat I}$ is the $2\times 2$ identity matrix and 
${\hat \tau}_3$ is the third Pauli matrix. 

At the second order in a gradient expansion 
\cite{loktev0,babaev,loktev,tempere} of $\Sigma_{\theta}$ 
the partition function eventually can be written as 
\beq 
{\cal Z} = \exp{\left\{ - {S_{mf}\over \hbar} \right\}} 
\int {\cal D}[\theta] \exp{\left\{- {S_{\theta}\over \hbar} \right\}} \; ,   
\label{cisiamo}
\eeq
where $S_{mf}$ is given by Eq. (\ref{omega-sp}), while 
the action functional $S_{\theta}$ of the phase is 
given by \cite{loktev0,babaev,loktev,tempere}
\beq 
S_{\theta} = \int_0^{\hbar\beta} d\tau \int_{L^2} d^2{\bf r} 
\left\{ {J\over 2} \ (\nabla\theta)^2 + {K_{\theta\theta}\over 2} 
\ (\partial_{\tau}\theta)^2 \right\} \; , 
\label{thetact} 
\eeq
where 
\beq
J = {\hbar^2\over 4 m{L^2}} \sum_{{\bf k}} 
\left[ 1 - {{\hbar^2k^2\over 2m} -\mu\over E_k}X_T(E_k) - 
{\hbar^2k^2\over 2m} X_T'(E_k)
\right] \; , 
\label{J}
\eeq
is the stiffness,  
\beq
K_{\theta\theta} = {\hbar^2\over 4 {L^2}} 
\sum_{{\bf k}} 
\left[ {\Delta_0^2\over E_k^3}X_T(E_k) + {({\hbar^2k^2\over 2m} -\mu)^2
\over E_k^2} X_T'(E_k) \right] \; .
\label{K} 
\eeq
is the phase susceptibility, and 
$X_T(E_k)=\tanh{(\beta E_k/2)}$. 
Notice that $J$ and $K_{\theta\theta}$ 
are non trivial functions of $T$, $\mu$ and $\Delta_0(T,\mu,\epsilon_B)$, 
and from Eqs. (\ref{gap-r}) and(\ref{number}) one gets 
$\Delta_0$ and $\mu$ as a function of $T$, $\epsilon_B$ and $n$. 

The action functional (\ref{thetact}) has the form 
of a 2D quantum XY model \cite{nagaosa,atland,stoof}, where
the Goldstone field $\theta({\bf r},\tau)$ 
is defined in principle as an angular 
variable. However, it is well known \cite{nagaosa,atland,stoof} 
that, in addition to the characteristic temperature $T^*$ below which 
quantized vortices develop, there is another relevant 
temperature in our system: 
the temperature $T_{BKT}$ of the Berezinskii-Kosterlitz-Thouless 
superfluid-normal phase transition,  
characterized by the binding
of quantized vortices below $T_{BKT}$.  
The contribution of vortices below $T_{BKT}$ then becomes 
irrelevant at large distance scales and the 
field $\theta$ loses its angular character, thus justifying a 
Gaussian treatment at small energy-momentum. 
This critical temperature $T_{BKT}$ can be 
estimated by solving self-consistently 
\cite{loktev0,babaev,tempere} :
\beq 
k_B \, T_{BKT} = {\pi\over 2} J(T_{BKT}) \; ,    
\label{tbkt}
\eeq
where $J(T)$ is defined by Eq. (\ref{J}) with $\mu$ and $\Delta_0$ 
given by the solutions of
the gap and number equations Eqs. (\ref{gap-r}) and(\ref{number}). 
Following the approach adopted by various authors 
\cite{marini,loktev0,babaev,loktev}, we use the lowest-order 
mean-field functions $\Delta_0$ and $\mu$ and 
plug them into the new (higher-order) effective action. 
Strictly speaking, instead of Eq. (\ref{number})  one should use
a modified number equation, where $\Omega_{mf}$ is substituted by 
$\Omega_{mf}+\Omega_{flu}$ with $\Omega_{flu}$ taking into account 
fluctuations \cite{tempere,tempere2}. However, at zero temperature 
$\Omega_{flu}$ reduces to the zero-point energy of a bosonic gas 
with excitations $c_s \hbar k$, and on the basis of dimensional 
regularization \cite{leibb} one can set $\Omega_{flu}=0$. 

\begin{figure}[t]
\centerline{\epsfig{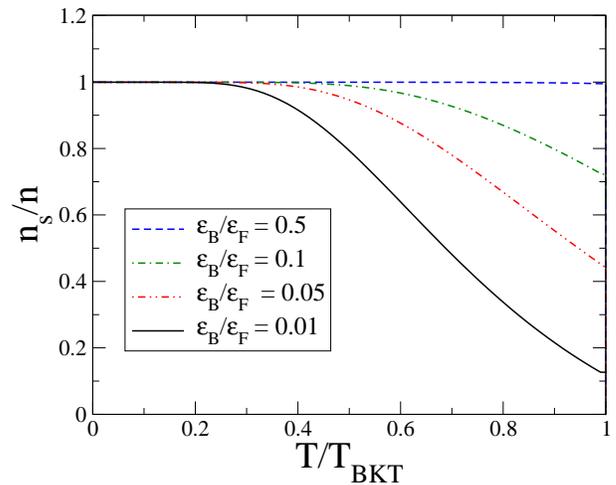}}
\small 
\caption{(Color online). Superfluid fraction $n_s/n$ 
as a function of the scaled temperature $T/T_{BKT}$ 
for different values of the scaled binding energy 
$\epsilon_B/\epsilon_F$, where $\epsilon_F=(\hbar^2/m)\pi n$ 
is the Fermi energy.} 
\label{fig2}
\end{figure}

The solid curve of Fig. \ref{fig1} shows $k_BT_{BKT}$ in units 
of the 2D Fermi energy $\epsilon_F$ as a function of the 
scaled binding energy $\epsilon_B/\epsilon_F$. The curve 
approaches very quickly its asymptotic value \cite{loktev0,babaev} 
\beq 
k_B T_{BKT} = {1\over 8} \epsilon_F \; . 
\eeq
The domain between two curves 
shown in Fig. \ref{fig2} is the so-called pseudo-gap region 
\cite{loktev0,babaev,loktev,tempere,tempere2} 
where vortices proliferate and a more careful 
treatment of $\theta$ as angular variable is needed, 
leading in particular to a gap for the Goldstone field. 

Since ${\bf v}_s= (\hbar/m) {\boldsymbol \nabla} \theta$ is 
the superfluid velocity, the term $(J/2)(\nabla \theta)^2$ 
may be identified with the superfluid kinetic energy density 
$(1/2)n_s v_s^2 $, where 
\beq
n_s = {4m\over \hbar^2} J 
\eeq 
is the superfluid number density. The 
renormalization group theory \cite{stoof} dictates that 
for a 2D uniform system above $T_{BKT}$ the phase stiffness $J$,  
and consequently also superfluid density $n_s$, is strictly zero. 
This result implies a jump to zero of the superfluid density at $T_{BKT}$ 
\cite{nagaosa,stoof}. In Fig. \ref{fig2} we report the 
superfluid fraction $n_s/n$ as a function of the scaled 
temperature $T/T_{BKT}$ for different values of the scaled 
binding energy $\epsilon_B/\epsilon_F$. The figure clearly shows that 
the superfluid fraction $n_s/n$ is equal to one at very low temperatures 
and decreases monotonically by increasing the temperature $T$. 
Moreover, for very small values of the scaled binding energy 
$\epsilon_B/\epsilon_F$ the superfluid fraction $n_s/n$ is quite small 
at $T=T_{BKT}$ while for larger values 
of the scaled binding energy $\epsilon_B/\epsilon_F$ the superfluid 
fraction $n_s/n$ remains close to one up to $T=T_{BKT}$. Notice 
that $\epsilon_B/\epsilon_F=0.5$ still corresponds to a positive 
zero-temperature chemical potential $\mu(0)$, i.e. to a system 
in the BCS regime. 

\section{Phase and amplitude fluctuations and sound velocity}

Any superfluid system admits a density wave, the so-called first sound, 
where the velocities of superfluid and normal components are in-phase 
\cite{nagaosa,atland}. The velocity of the Goldstone mode 
is nothing else than the first 
sound velocity of the superfluid \cite{nagaosa,atland} and it is given by 
\beq 
c_s=\sqrt{J\over K} \; , 
\label{tanata}
\eeq 
where $J$ is the stiffness and $K$ is the susceptibility. 
Within the phase-only approach of the previous section we 
have $K=K_{\theta\theta}$, and using Eqs. (\ref{J}) and (\ref{K}) 
at zero temperature one immediately finds 
\beq
J={\epsilon_F\over 4\pi} \; , 
\eeq
and 
\beq 
K_{\theta\theta}={m\over 4 \pi} {\epsilon_F\over \epsilon_F
+{1\over 2}\epsilon_B} \; , 
\label{zuzana}
\eeq 
and consequently, using Eq. (\ref{tanata}) 
with $K=K_{\theta\theta}$, we obtain 
\beq
c_{s}={v_F\over \sqrt{2}} 
\sqrt{1+{1\over 2}{\epsilon_B\over \epsilon_F}} 
\quad\quad \mbox{at $T=0$ (phase-only)} \; ,  
\eeq
where $v_F=\sqrt{2\epsilon_F/m}$ is the Fermi velocity 
and $\epsilon_F=(\hbar^2/m)\pi n$ is the Fermi energy. 
We stress that this result is obtained by completely neglecting 
amplitude fluctuations $\sigma({\bf r},\tau)$
of the order parameter $\Delta({\bf r},\tau)$.  

Recently Schakel \cite{schakel} has analyzed the 3D BCS-BEC crossover 
at zero temperature considering both phase $\theta({\bf r},\tau)$ 
and amplitude $\sigma({\bf r},\tau)$ fluctuations 
in $\Delta({\bf r},\tau)$. Following the procedure of 
Schakel \cite{schakel}, in our zero-temperature 2D system 
after integration over $\sigma({\bf r},\tau)$  
we obtain the action functional $S_{\theta}$ of Eq. (\ref{thetact})
with the stiffness $J$ still given by Eq. (\ref{J}) but 
with a new $K$ instead of $K_{\theta\theta}$. In particular, 
the new susceptibility $K$ is given by 
\beq 
K = {K_{\theta\theta} K_{\sigma\sigma} - K_{\sigma\theta}^2 
\over K_{\sigma\sigma}} \; , 
\eeq
which is a non trivial combination of the phase-only susceptibility 
$K_{\theta\theta}$ given by Eq. (\ref{K}), the amplitude-only 
susceptibility $K_{\sigma\sigma}$ 
and the amplitude-phase susceptibility $K_{\sigma\theta}$. 
Note that only when amplitude and phase fluctuations are decoupled, 
i.e. when $K_{\sigma\theta}\simeq 0$ one 
obtains $K\simeq K_{\theta\theta}$. 

\begin{figure}[t]
\centerline{\epsfig{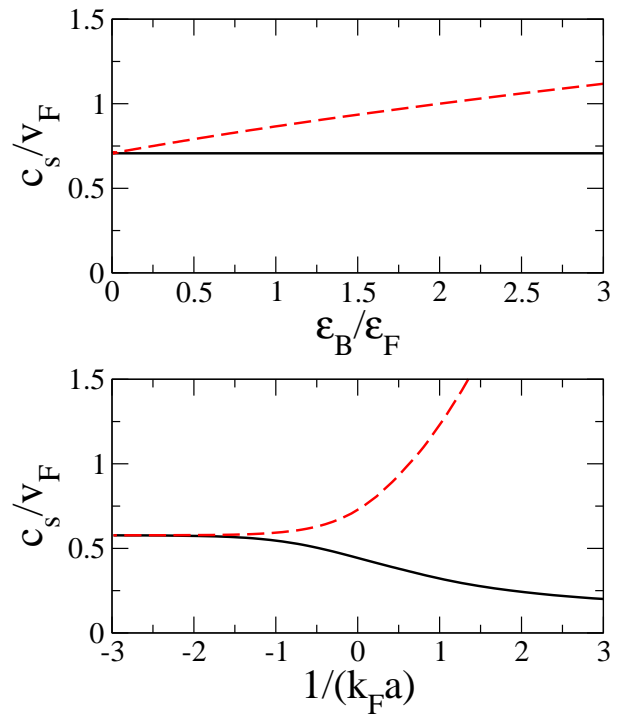}}
\small 
\caption{(Color online). Sound velocity $c_s$ at zero temperature ($T=0$)
taking into account only phase fluctuations (dashed lines) or 
both phase and amplitude fluctuations (solid lines) of the order 
parameter. Upper panel: 
2D scaled sound velocity $c_{s}/v_F$ as a function of the scaled 
binding energy $\epsilon_B/\epsilon_F$ of the 2D Fermi superfluid. 
Lower panel: 3D scaled sound velocity $c_{s}/v_F$ as a function of the scaled 
inverse interaction strength $1/(k_Fa)$ of the 3D Fermi superfluid 
with scattering length $a$. 
Here $\epsilon_F=\hbar^2k_F^2/(2m)$ is the Fermi energy and 
$v_F=\sqrt{2\epsilon_F/2}$ the Fermi velocity.} 
\label{fig3}
\end{figure} 

At zero temperature, we get (see also \cite{schakel}) 
the following formulas 
\beqa
K_{\theta\theta} &=& - {\hbar^2\over 4L^2} 
\left({\partial^2 \Omega_{mf}\over \partial \mu^2}\right)_{L^2,T=0} \; , 
\\
K_{\sigma\sigma} &=& - {\hbar^2\over 4L^2} 
\left({\partial^2 \Omega_{mf}\over \partial \Delta_0^2}\right)_{L^2,T=0} \; , 
\\
K_{\sigma\theta} &=&  {\hbar^2\over 4L^2} 
\left({\partial^2 \Omega_{mf}\over \partial 
\Delta_0\partial \mu}\right)_{L^2,T=0} \; .   
\eeqa 
By using these formulas for our 2D superfluid system 
we easily find that $K_{\theta\theta}$ is indeed 
given by Eq. (\ref{zuzana}), 
while $K_{\sigma\sigma}$ and $K_{\sigma\theta}$ are 
\beqa
K_{\sigma\sigma} &=& - {m\over 8\pi \epsilon_B} {\Delta_0^2\over 
\epsilon_F +{1\over 2}\epsilon_B} \; , 
\\
K_{\sigma\theta} &=& {m\over 8\pi} {\Delta_0\over 
\epsilon_F +{1\over 2}\epsilon_B} \; .
\eeqa
It follows that the sound velocity of the 2D superfluid system reads 
\beq 
c_s = {v_F\over \sqrt{2} } \quad\quad 
\mbox{at $T=0$ (phase and amplitude)} \; ,  
\label{kafone}
\eeq
which is exactly the 2D result obtained some years ago 
by Marini, Pistolesi and Strinati \cite{marini}. 
Taking into account both phase and amplitude fluctuations of the 
order parameter (Gaussian fluctuations), at zero temperature 
the 2D sound velocity $c_s$ does not depend on the binding 
energy $\epsilon_B$ of pairs. 

Thus, as reported in the upper panel of Fig. \ref{fig3}, taking into 
account only phase fluctuations of the order parameter leads to a quite 
different behaviour of the zero temperature speed of sound in the 2D Fermi 
superfluid from that obtained by considering both phase and amplitude 
fluctuations.
While the latter does not depend on $\epsilon_B$, the former increases 
with it and diverges in the deep BEC regime. 
A similar behaviour is obtained in 3D for the dependence of the speed of 
sound on the the scaled inverse interaction strength $1/(k_Fa)$ 
which we report for completeness in the lower panel of Fig. \ref{fig3}. 
Also this panel shows that only in the deep BCS regime,  
where $1/(k_Fa)\ll -1$, the two approaches 
give the same results $c_s\simeq v_F/\sqrt{3}$ while, again, 
the phase-only sound velocity diverges 
in the BEC regime. 

We now show that the Gaussian (phase plus amplitude) 
result, Eq. (\ref{kafone}), can be re-derived 
by using simple thermodynamic relations \cite{combescot} 
and it can also be easily extended 
at finite temperature. In fact, 
according to Landau \cite{landau} and Kalatnikov \cite{khala} 
the first sound velocity $c_s$ is given by
\beq 
m \, c_s^2 = \left({\partial P \over \partial n}\right)_{L^2,{\bar S}} \; , 
\eeq
where $P$ is the pressure and 
$\bar{S} = S/N$ is the entropy per particle of the superfluid. Moreover, 
at zero temperature it holds the following equality 
\beq 
\left({\partial P\over \partial n}\right)_{L^2,0} 
=n \left( {\partial \mu\over \partial n} \right)_{L^2} \; . 
\label{forse}
\eeq
Using Eq. (\ref{ita01}) we immediately obtain Eq. (\ref{kafone}). 

At finite temperature we can determine the sound velocity $c_s$ using 
the elegant formula of thermodynamics 
\beq 
m \, c_s^2 \simeq n \left( {\partial \mu\over \partial n} \right)_{L^2,T} \; . 
\eeq
Numerically we find that $c_s$ remains 
close to $1/\sqrt{2}$ for any temperature $T$ 
(up to $T_{BKT}$) and for any value of the scaled 
binding energy $\epsilon_B/\epsilon_F$. This is in full agreement 
with experiments with 3D superfluids 
like $^{4}$He liquid and unitary Fermi gas 
the sound velocity $c_s$ does not depend 
significantly on the temperature $T$. 

\begin{figure}[t]
\centerline{\epsfig{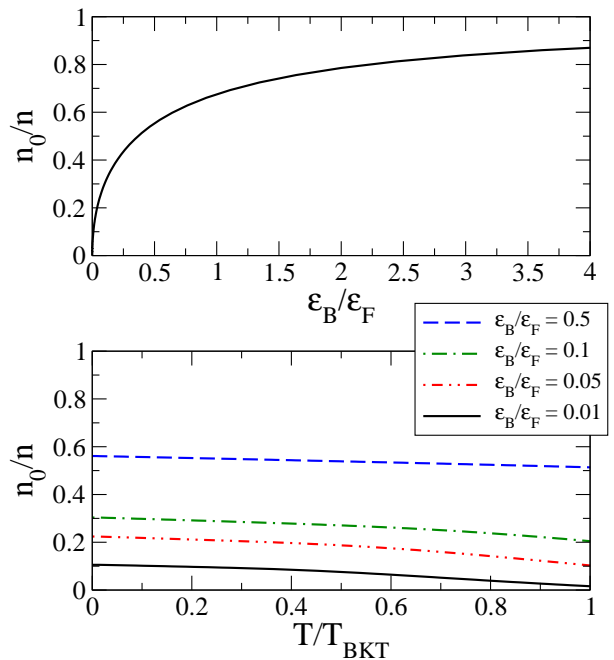}}
\small 
\caption{(Color online). Upper panel: zero-temperature condensate 
fraction $n_0/n$ vs scaled binding energy $\epsilon_B/\epsilon_F$. 
Lower panel: quasi-condensate fraction $n_0/n$ as a fuction 
of the scaled temperature $T/T_{BKT}$ for different values 
of the scaled binding energy $\epsilon_B/\epsilon_F$.} 
\label{fig4}
\end{figure} 

\section{Algebraic long-range order and quasi-condensate fraction}

As previously discussed, according to the Mermin-Wagner-Hohenberg-Coleman 
theorem \cite{mermin,hohenberg,coleman}, 
in a 2D uniform quantum system of interacting identical particles 
one can find true condensation, i.e off-diagonal-long-range-order 
(ODLRO), only at zero temperature ($T = 0$). Instead, the system can have
quasi condensation, i.e. algebraic-long-range-order (ALRO), below a
critical finite temperature that is usually identified 
with the Berezinskii-Kosterlitz-Thouless temperature $T_{BKT}$ 
\cite{nagaosa,stoof}. 
In the case of our 2D Fermi system the two-body density matrix 
\beq 
\rho_2({\bf r}_1,{\bf r}_2,{\bf r}_3,{\bf r}_4) = 
\langle 
\bar{\psi}_{\uparrow}({\bf r}_1,0) \, \bar{\psi}_{\downarrow}({\bf r}_2,0)
\, \psi_{\downarrow}({\bf r}_3,0) \, \psi_{\uparrow}({\bf r}_4,0)
\rangle
\eeq
shows ODLRO at $T=0$ \cite{sala} and ALRO 
for $0<T<T_{BKT}$. In particular, by using Eq. (\ref{chi}) 
and introducing the center-of-mass positions of the two Cooper pairs, 
given by ${\bf R}=({\bf r}_1+{\bf r}_2)/2$ and 
${\bf R}'=({\bf r}_3+{\bf r}_4)/2$, and their relative distances 
${\bf r}={\bf r}_2-{\bf r}_1$ and ${\bf r}'={\bf r}_4-{\bf r}_3$, 
for $|{\bf R}-{\bf R}'|\to \infty$ we can write 
\beqa
\rho_2({\bf r}_1,{\bf r}_2,{\bf r}_3,{\bf r}_4) 
&\simeq& 
F^*({\bf r}) \, F({\bf r}') \, 
\langle
e^{i\left( \theta({\bf R},0) - \theta({\bf R}',0) \right)} 
\rangle 
\nonumber
\\
&\simeq& 
F^*({\bf r}) \, F({\bf r}') \, 
e^{ -{1\over 2} \langle \left( \theta({\bf R},0) - \theta({\bf R}',0) 
\right)^2 \rangle} 
\nonumber
\\
&\simeq& 
F^*({\bf r}) \, F({\bf r}') \, 
\left( {R_{0}\over |{\bf R}-{\bf R}'|} \right)^{k_BT\over 8\pi J} 
\label{secicredi}
\eeqa
where $R_{0}=2c_s/(k_BT)$ is the coherence length scale of 
phase fluctuations \cite{stoof,loktev} and 
\beqa 
F({\bf r}') &=& 
\langle \chi_{\downarrow}({\bf r}_3,0) \chi_{\uparrow}({\bf r}_4,0) 
\rangle 
\nonumber
\\
&=& {1\over L^2} 
\sum_{\bf k} {\Delta_0\over 2E_k} 
\tanh(\beta E_k/2) \, e^{i{\bf k}\cdot {\bf r}'}
\eeqa
is the mean-field wavefunction of the Cooper 
pair \cite{sala-odlro,ortiz,ohashi2}, such that 
\beq 
n_0 = 2 \int d^2{\bf r}' \, |F({\bf r}')|^2 = 
{\Delta_0^2\over 2L^2} 
\sum_{\bf k} {\tanh^2(\beta E_k/2)\over E_k^2} \
\eeq
is the quasi-condensate density of atoms in the 2D superfluid. 

At $T=0$ Eq (\ref{secicredi}) displays ODLRO, i.e there is no algebraic 
decay of the off-diagonal part of the two-body density matrix, 
and $n_0$ is the true condensate density of 
the system (see also \cite{sala}). 
In the upper panel of Fig. \ref{fig4} we plot the zero-temperature 
condensate fraction $n_0/n$ as a function of the 
scaled binding energy $\epsilon_B/\epsilon_F$. 
At finite temperature Eq (\ref{secicredi}) displays ALRO, i.e. there 
is algebraic decay of the off-diagonal part of 
the two-body density matrix, and $n_0$ is the quasi-condensate density of 
the system (see \cite{stoof} for the bosonic case). 
In the lower panel of Fig. \ref{fig4} we plot the 
quasi-condensate fraction $n_0/n$ as a fuction 
of the scaled temperature $T/T_{BKT}$ for different values 
of the scaled binding energy $\epsilon_B/\epsilon_F$. 
The figure clearly shows that for large values of the scaled 
binding energy $\epsilon_B/\epsilon_F$ the quasi-condensate fraction 
$n_0/n$ is practically independent on the temperature up to the 
Berezinskii-Kosterlitz-Thouless critical temperature $T_{BKT}$. 

\section{Conclusions}

By using the path integral formalism and the thermodynamics 
of superfluids we have calculated the superfluid density,  
the sound velocity, and the quasi-condensate density 
of a 2D superfluid made of ultracold alkali-metal atoms 
in the BCS-BEC crossover. We have considered 
both phase and amplitude fluctuations 
of the order parameter showing that amplitude fluctuations 
are necessary to recover within the path integral formalism 
the sound velocity one gets alternatively from the 
mean-field equation of state by using familiar 
thermodynamics relationships. 
Our results are obtained below the critical 
temperature $T_{BKT}$ of the Berezinskii-Kosterlitz-Thouless 
phase transition, where there is quasi-condensation and the 
Goldstone field of phase fluctuations is still massless. 
Notice that the crucial role of phase fluctuations 
on the Berezinskii-Kosterlitz-Thouless transition 
has been very recently investigated 
with the attractive Hubbard model by Erez and Meir \cite{erez}. 
We believe that a reliable description 
of the pseudo-gap region above $T_{BKT}$ \cite{nature}, 
where the Goldstone field of phase fluctuations becomes 
gapped with exponential decay of correlations, requires a more 
sophisticated self-consistent approach 
to the phase fluctuations \cite{marchetti}. We are currently  
working on this issue. 

\section*{Acknowledgments}

The authors thank Luca Dell'Anna, Adriaan Schakel, Sergei Sharapov, 
and Jacques Tempere for useful discussions and suggestions. 
The authors acknowledge for partial support 
Universit\`a di Padova (Research Project 
"Quantum Information with Ultracold Atoms in Optical Lattices"), 
Cariparo Foundation 
(Excellence Project "Macroscopic Quantum Properties 
of Ultracold Atoms under Optical Confinement"), 
and Ministero Istruzione Universita Ricerca 
(PRIN Project "Collective Quantum Phenomena: from 
Strongly-Correlated Systems to Quantum Simulators").

\end{document}